\documentclass[twocolumn,nofootinbib]{revtex4}

\usepackage{ORI_Group_style}
\usepackage[normalem]{ulem}
\usepackage{float}
\usepackage[caption=false]{subfig}

\usepackage{footnote}

\begin{document}

\title{Repulsive vacuum-induced forces on a magnetic particle}

\author{Kanupriya Sinha}
\email{kanupriyasinha@gmail.com}
\affiliation{Institute for Theoretical Physics, University of Innsbruck, A-6020 Innsbruck, Austria.}
\affiliation{Institute for Quantum Optics and Quantum Information of the
Austrian Academy of Sciences, A-6020 Innsbruck, Austria.}

\begin{abstract}
We study the possibility of obtaining a repulsive vacuum-induced force for a magnetic point particle near a surface. Considering the toy model of a particle with an electric-dipole transition and a large magnetic spin, we analyze the interplay between the repulsive magnetic-dipole   and the attractive electric-dipole  contributions  to  the total Casimir-Polder  force. Particularly noting that the magnetic-dipole interaction is  longer-ranged than the electric-dipole  due to the difference in their respective characteristic transition frequencies, we find a regime where the repulsive magnetic contribution to the total force can potentially exceed the attractive electric part in magnitude for a sufficiently large spin. We analyze ways to further enhance the magnitude of the repulsive magnetic Casimir-Polder force for an excited particle, such as by preparing it in a ``super-radiant''  magnetic sub-level, and designing surface resonances close to the magnetic transition frequency.
\end{abstract}

\maketitle

\global\long\def\eqn#1{\begin{align}#1\end{align}}
\global\long\def\ket#1{\left|#1\right\rangle }
\global\long\def\bra#1{\left\langle #1\right|}
\global\long\def\bkt#1{\left(#1\right)}
\global\long\def\sbkt#1{\left[#1\right]}
\global\long\def\cbkt#1{\left\{#1\right\}}
\global\long\def\abs#1{\left\vert#1\right\vert}
\global\long\def\der#1#2{\frac{{d}#1}{{d}#2}}
\global\long\def\re{\mathrm{Re}}
\global\long\def\im{\mathrm{Im}}
\global\long\def\dd{\mathrm{d}}
\global\long\def\ddd{\mathcal{D}}

\global\long\def\avg#1{\left\langle #1 \right\rangle}
\global\long\def\mr#1{\mathrm{#1}}
\global\long\def\mb#1{{\mathbf #1}}
\global\long\def\mc#1{\mathcal{#1}}
\global\long\def\bs#1{\boldsymbol{#1}}
\global\long\def\tr{\mathrm{Tr}}
\global\long\def\dbar#1{\Bar{\Bar{#1}}}
\global\long\def\cev#1{\overleftarrow{#1}}
\global\long\def\vec#1{\overrightarrow{#1}}
\global\long\def\non{\nonumber}
\section{Introduction}
That the quantum fluctuations of the electromagnetic (EM) field in vacuum state can lead to forces between neutral objects
is a fascinating feature of quantum electrodynamics (QED) \cite{Milonni}. Such fluctuation forces, often addressed by different names depending on the geometry, separation and material properties of the interacting objects, such as van der Waals  \cite{vdW}, London  \cite{London}, Casimir-Polder  \cite{CP1948}, Casimir-Lifshitz  \cite{Lifshitz}, or more generally Casimir \cite{Casimir1948} forces, arise as a result of the interaction mediated between the fluctuating dipole moments that constitute two neutral
bodies via the quantum  fluctuations of the EM field.

When considering atom-surface interactions,  Casimir-Polder (CP) forces become significant in comparison to  externally applied forces at distances smaller than atomic wavelengths. Being typically attractive and short-ranged,  fluctuation forces are considered as a detrimental influence when trying to trap and control quantum systems near surfaces. For example, the typical magnetic and optical trap forces are easily overcome by vacuum forces at the nanoscale. As a result, when trying to interface trapped atoms with surfaces, the atoms tend to be lost from the relatively weak trapping potentials and adhere to surfaces \cite{Vuletic}.

With growing efforts towards miniaturization of photonic systems both with the fundamental motivation to explore
quantum phenomena at increasingly shorter length scales, and the practical goal of replacing large-scale optical
elements with modular on-chip architectures \cite{Aidan2013,Treutlein2007,AtomChipGravimeter,HarocheChip,Hinds,Folman2016,Meek2009}, atom-surface interactions have become an increasingly relevant
aspect of understanding and designing nanoscale photonic devices.  For example, Casimir interactions become an inevitable element of consideration in trapping schemes \cite{Darrick2013,spinflip3}, 
surface-modification of decay rates \cite{SpEm1,SpEm2,Pablo2017,Asenjo2017}, and decoherence of atomic spins \cite{spinflip1, spinflip2, superchip} when trapping atoms near nanofibers \cite{fiber1, fiber2,fiberKimble,fiberLuis,Balykin2004}, photonic crystals \cite{PhC1,PhCLukin,PhCKimble}, micro- and nano-scale cavities \cite{Alton2011,Aoki2014}. Thus, given that vacuum forces play an important role in state-of-the-art experiments, it is  interesting to explore whether they can be engineered in a way to achieve better control and coherence of quantum systems interacting at nanoscales. Particularly, we examine here the possibility of using repulsive vacuum forces to stably trap a particle near a surface.

However, an analog of the Earnshaw's theorem for fluctuation forces forbids stable equilibria for non-magnetic objects separated by vacuum \cite{nogo}.  Some possible  ways to overcome this no-go theorem are \cite{Prachi}:  going out-of-equilibrium using temperature gradients \cite{thermal1,thermal2,thermal3} or external drives \cite{KSDEC}, replacing the vacuum by a medium with appropriate permittivity  relative to the interacting objects \cite{Munday1,Sabisky,Ishikawa}, using material anisotropies \cite{Munday2017,Diego2008,DiegoRev} and topological properties of the interacting bodies \cite{Justin<3, PabloR2014}, or designing specific geometrical configurations \cite{Levin2010}.

As another way to circumvent  the no-go theorem, one can  use the  magnetic response of the interacting bodies \cite{Boyer,Klich,Genov}. For example,  when considering the force on a magnetic atom interacting
with the vacuum EM field near a perfectly conducting surface, it is known that the electric-dipole (ED) interaction between the atom and the EM field leads to an attractive CP force, while the magnetic-dipole (MD) interaction leads to a repulsive force \cite{Boyer1969, Datta1981}. Previous works that have studied repulsive CP forces due to MD interaction between atoms and surfaces  \cite{Carsten2005, Skagerstam,Haakh2009} show that  the MD interaction induced repulsion is limited due to the smallness of magnetic interaction in comparison to the electric.  

Keeping this in consideration, in this paper we study the possibility of using the  magnetic-dipole interaction between a point-like magnetic particle with a large spin and a surface to realize an overall repulsive CP force. We find that the force due to the MD interaction has two components -- a braodband Casimir-Polder contribution, and a zero frequency contribution coming from the magnetostatic interaction between the magnetic dipole and its image, which can be significant  for a perfect conductor-like surface. We show that for a large enough spin, in the appropriate distance regime where one has a weak short-ranged ED contribution while the MD contribution is considerable, one can have an overall repulsive vacuum-induced force. We then study two particular ways of further enhancing the magnetic CP contribution for a particle in an excited magnetic sublevel -- using “superradiance”-like effects \cite{CollectiveCP}, and
by engineering the response of the surface at the resonant
frequency for the MD transition \cite{KSDEC}.

The paper is structured as follows. In section \ref{model}, we  present a theoretical model to describe a magnetic point particle interacting with the vacuum EM field in the presence of a surface. We study the surface-induced modifications to the internal dynamics of the atom as described by the second order Born-Markov master equation in section \ref{BMME}. In  section \ref{groundPC}, we analyze the force on a ground state  particle  with an arbitrarily large spin near a perfectly conducting surface coming from  ED and MD interactions. Further, adding gravity, we study the feasibility of creating stable equilibrium by combining a repulsive magnetic force and the attractive gravitational force. In section \ref{differentsurfaces}, we consider the particle near  a metal surface described by Drude and plasma models.  In section \ref{CPexcited}, we study potential ways to preferentially enhance the repulsive magnetic CP force relative to the attractive electric CP force by considering the particle to be in an excited magnetic sublevel. We discuss the conclusions and prospects of our work in section \ref{discussion}.

\section{Model}
\label{model}

We consider a fixed magnetic point particle at the position $\mb{r}_0 = ( 0,0,z_0)^T$ $(z_0>0)$ placed above a planar  medium that occupies the  half-space $z<0$. The half-space around the particle $(z>0)$ is assumed to be  vacuum. The internal degrees of the particle consist of an electric-dipole transition between ground and excited states $\ket{a}$ and $\ket{b}$, respectively, and a magnetic spin  of dimension $S$. We assume a classical magnetic field $\mb{B}_{0}(\mb{r}_0) = \abs{\mb{B}_0(\mb{r}_0)}\mb{e}_z$ along the $z$-axis at the position of the particle. We consider the particle to be in a magnetic state $\ket{S, m_S}$, such that $\hat{\mb{S}}^2 \ket{S,m_S} = S(S+1) \ket{S,m_S}$, and ${\hat{S}_z}\ket{S,m_S} = m_S \ket{S,m_S},$ with $m_S\in\{-S,\ldots,S-1,+S\}$. We note  that the spin operators $(\hat{ S}_x, \hat{ S}_y, \hat{ S}_z)$ obey the canonical commutation relations $[\hat S_i,\hat S_j] = \text{i} \epsilon_{ijk} \hat S_k$.

To describe the interaction between the particle and the vacuum EM field in the presence of a surface, we write the total Hamiltonian as \eqn{\hat H_{\mr{tot}} = \hat H_{P} +\hat H_F +\hat H_{PF}.}
The first term
\eqn{\hat H_P = \hbar\omega_e \hat\sigma_{+}\hat\sigma_{-} + \hbar \omega_m \hat{{S}}_z,}
refers to the free Hamiltonian of the particle, with $\omega_{e}$ as the transition frequency for the ED transition. The ladder operators for the ED transition are defined as $\hat\sigma_+ \equiv \ket{b}\bra{a}\otimes \mathbb{1} = (\hat\sigma_-)^\dagger$. The Zeeman splitting $\omega_m = \gamma_0 |\mb{B}_0 (\mb{r}_0)|$ is given by the MD interaction term $- \hat{\bf m} \cdot  \mb{B}_{0}(\mb{r}_0) = \hbar \omega_m  \hat{ {S}}_z,$  where  $\hat{\mb{m}} = -\hbar \gamma_0 \hat{\mb{S}}$ corresponds to the magnetic moment operator and $\gamma_0$ is the gyromagnetic ratio \cite{Sakurai}.  The Hamiltonian $\hat H_F$ describing the free dynamics of the  vacuum EM field in the presence of the surface is defined in Eq.\,\eqref{hv} \cite{Buh1, Buh2}.

The interaction between the particle and the vacuum EM field in the multipolar coupling scheme,  using the electric and magnetic dipole approximations, is given as \cite{CCT1}
\eqn{\hat 
H_{PF} =& -\hat{\mb{d}}\cdot\hat{\mb{E}}\bkt{\mb{r}_0} -\hat{\mb{m}}\cdot\hat{\mb{B}}\bkt{\mb{r}_0},\non\\
=& - \bkt{\mb{d}\hat\sigma_+ +\mb{d}^\ast\hat\sigma_-} \cdot\hat{\mb{E}}\bkt{\mb{r}_0}\non\\
&+\hbar \gamma_0\left[ \hat{S}_z \mb{e}_z+\hat{{S}}_+  \mb{e}_+ +  \hat{{S}}_-  \mb{e}_- \right] \cdot\hat{\mb{B}}\bkt{\mb{r}_0}.
}
 Here  $\mb{d} \equiv \sqrt{2} \vert\langle b\vert\hat{\mb{d}} \vert a\rangle\vert \mb{e}_+ $,  $\mb{e}_\pm\equiv (\mb{e}_x \mp i \mb{e}_y)/{2}$, and  the spin raising and lowering operators $\hat{S}_\pm\equiv \hat{ S}_x\pm i\hat{ S}_y$ are defined as
\eqn{ \label{Spm}\hat{S}_\pm \ket{S, m_S } = \sqrt{S\bkt{S+1} - m_S \bkt{m_S\pm 1}}\ket{S, m_S\pm 1}. }

We have assumed the ED moment to be circularly polarized for later convenience.  The electric  and magnetic field  operators for  the medium-assisted EM field, evaluated at the position of the particle,  $\hat{\mb{E}}(\mb{r}_0)$ and $\hat{\mb{B}}(\mb{r}_0)$,  are defined in Eq.\,\eqref{Era} and Eq.\,\eqref{Bra} respectively.

 
 \section{Surface-induced modifications to the internal dynamics of the particle}
 \label{BMME}
To find the influence of the medium-assisted EM field on the particle, we derive the surface-induced modifications to the  master equation describing the dynamics of the reduced density matrix $\hat \rho_P$, that corresponds  to the internal degrees of freedom of the particle \cite{KSDEC}. Assuming that the particle and the field are weakly coupled, and that the  EM field bath correlations decay much faster than the relaxation time scale for the particle's internal dynamics, we use the Born and Markov approximations to write the equation of motion for $\hat \rho_P$ as  \cite{BPbook}
\eqn{
&\der{\hat \rho_P}{t} = \non\\ &-\frac{1}{\hbar^2} \tr_F \int_0^\infty \dd\tau\sbkt{\tilde H^{\mr{sc}}_{PF}\bkt{t}, \sbkt{\tilde H^{\mr{sc}}_{PF}\bkt{t-\tau}, \hat \rho_P (t)\otimes\hat \rho_F}},
}
where  $\tilde{{H}}^{\mr{sc}}_{PF}(t)\equiv e^{-i\bkt{\hat H_P +\hat H_F}t}{\hat {H}}^{\mr{sc}}_{PF}e^{i\bkt{\hat H_P +\hat H_F}t} $ stands for the interaction Hamiltonian in the  interaction picture, including only  the part of the EM field scattered off the surface, as in  Eq.\,\eqref{havsc}. The reduced density  matrix $\hat \rho_F = \ket{0}\bra{0}$ refers to that of the vacuum EM field.  Tracing out the field, we obtain the surface-modifications to the second-order Born-Markov master equation for the particle dynamics as
\eqn{
\der{\hat\rho_P}{t}  = -\frac{i}{\hbar} \sbkt{ \Delta\hat H_P, \hat \rho_P} + \Delta\mathcal{L}_P  \sbkt{\hat \rho_P},
}
where $\Delta\hat H_P $ corresponds to the surface-induced dispersive corrections to the particle Hamiltonian, and $\Delta\mathcal{L}_P$ refers to the surface-induced modifications to the dissipative dynamics of the particle. We remark  that we have considered here only the  part of the EM field scattered off the surface, without including the free-space contribution that leads to the Lamb shift and the free-space dissipation.

The corrections to the particle Hamiltonian are given by,
\eqn{\label{dHa} \Delta\hat H_P = & u_e^{(+)}\hat{\sigma}_+\hat{\sigma}_- + u_e^{(-)}\hat{\sigma}_-\hat{\sigma}_+\non\\&+ u_m^{(+)}\hat{S}_+\hat{S}_- + u_m^{(-)}\hat{S}_-\hat{S}_+ + u_m^{(z)} {\hat{{S}}_z}^2,
} The energy level shifts  $ u_e^{(-)}$  $ (u_e^{(+)})$ in Eq.\,\eqref{dHa}  to  the ground (excited) state coming from the ED interaction between the particle and the field are defined as \cite{CP1948, Buh1, Buh2, WylieSipe1, WylieSipe2},
\vspace{-0.2 in}
\begin{widetext}
\eqn{\label{de-}
 u_e^{(-)} =& \frac{2\mu_0 \omega_e \abs{\mb{d}}^2 }{\pi} \int_0^\infty \dd\xi \,\frac{\xi^2 }{\xi^2 +\omega_e^2} {\bkt{\mb{e}_-}^T\dbar G_{\text{sc}}\bkt{\mb{r}_0,\mb{r}_0, i\xi } \mb{e}_+},\\
\label{de+}
  u_e^{(+)} =& -\frac{2\mu_0 \omega_e\abs{\mb{d}}^2}{\pi} \int_0^\infty \dd\xi \,\frac{\xi^2 }{\xi^2 +\omega_e^2} \bkt{\mb{e}_+}^T\dbar G_{\text{sc}}\bkt{\mb{r}_0,\mb{r}_0, i\xi } \mb{e}_-- 2\mu_0 \omega_e^2\abs{\mb{d}}^2 \re\sbkt{\bkt{\mb{e}_+}^T\dbar G_{\text{sc}}\bkt{\mb{r}_0,\mb{r}_0,\omega_e }\mb{e}_-}.}
\end{widetext}
Here $\dbar{G}_{\mr{sc}}(\mb{r}_0,\mb{r}_0,\omega)$, given by Eq.\,\eqref{green}, refers to the scattering Green's tensor corresponding to a point dipole at $\mb{r}_0$ radiating near a surface \cite{GreenWelsch}.

Next, we consider  the shifts from the MD interaction in Eq.\,\eqref{dHa}. The energy level shifts $  u_{m}^{(-)}$ ($  u_m^{( +)}$) to the magnetic sublevels coming from the transition to the upper (lower) level, and $  u_m^{(z)}$ coming from the diagonal $(\sim \hat{{S}}_z)$ interaction term can be found as  follows \cite{Buh1, Buh2,  Skagerstam},
\vspace{-0.1 in}
\begin{widetext}
\eqn{
\label{dm-}
   u_m^{(-)}=& \frac{\hbar^2\mu_0 \omega_m\gamma_0^2}{\pi} \int_0^\infty \frac{\dd\xi}{\xi^2+\omega_m^2}{\bkt{{\bf e}_-}^T\dbar{\mc{G}}_{\mr{sc}}\bkt{\mb{r}_0,\mb{r}_0,i\xi}{\bf e}_+},\\
\label{dm+}
   u_m^{(+)} =&-\frac{\hbar^2\mu_0\gamma_0^2\omega_m }{\pi}\int_0^\infty  \frac{\dd\xi}{\xi^2+\omega_m^2}{\bkt{{\bf e}_+}^T \dbar{\mc{G}}_{\mr{sc}}\bkt{\mb{r}_0,\mb{r}_0,i\xi}{\bf e}_-}+\mu_0 \gamma_0^2\re\sbkt{\bkt{{\bf e}_+}^T \dbar{\mc{G}}_{\mr{sc}}\bkt{\mb{r}_0,\mb{r}_0,\omega_m}{\bf e}_-},\\
 \label{dmz}
    u_m^{(z)}=& \frac{\hbar^2\mu_0\gamma_0^2}{2} \lim_{\xi\rightarrow0}\sbkt{{\bkt{{\bf e}_z}^T \dbar{\mc{G}}_{\mr{sc}}\bkt{\mb{r}_0,\mb{r}_0,i\xi} {\bf e}_z}},
}
\end{widetext}

where we have defined the double curl of the scattering Green's tensor, given by  Eq.\,\eqref{greencc}, as \eqn{&\sbkt{\dbar{\mc{G}}_{\mr{sc}}\bkt{\mb{r}_0,\mb{r}_0,\omega}}_{il}\non\\
&\equiv\lim_{\mb{r}_1,\mb{r}_2\rightarrow\mb{r}_0}\epsilon_{ijk}  \epsilon_{nml} \frac{\partial^2}{\partial r_{1j} \partial r_{2m}} \sbkt{\dbar{G}_{\mr{sc}}\bkt{\mb{r}_1,\mb{r}_2,\omega}}_{kn}.}
The ground state shifts Eq.\,\eqref{de-} and Eq.\,\eqref{dm-}, coming purely from the energy non-conserving   terms in the particle-field interaction Hamiltonian,  can be  attributed to  the emission and reabsorption of a virtual photon scattered off the surface by the particle.   The energy conserving terms  lead to the excited level shifts as in Eq.\,\eqref{de+} and Eq.\,\eqref{dm+}. There are two contributions to the total excited state shifts, the first terms in Eq.\,\eqref{de+} and Eq.\,\eqref{dm+} correspond to the off-resonant process as in the ground state shifts, wherein the particle transits from the excited to the ground state and back, emitting and reabsorbing a virtual photon at a frequency $\omega\neq \omega_{e,m}$. In addition, the second term in Eq.\,\eqref{de+} and Eq.\,\eqref{dm+} represents  a resonant contribution that corresponds to the interaction of the particle with a real photon emitted at the resonant frequency for the ED or MD transitions. 

Additionally, we note that for the MD interaction, the diagonal part of the interaction Hamiltonian that goes as $\sim \hat S_z$ yields an energy shift  $  u_m ^{(z)}$, as given in Eq.\,\eqref{dmz}. Such a term is non-vanishing only for a surface with a non-zero response at static frequencies, for example, a perfect conductor. This contribution can be thus understood in terms of a static component of the magnetic dipole interacting with its image such that the normal component of the B field at the surface vanishes \cite{Barton, Griffiths, saslow}. We consider this static contribution to  the magnetic level shift as separate from the fluctuation-induced CP potentials in Eq.\,\eqref{de-}--Eq.\,\eqref{dm+}.


Considering the Liouvillian part of the master equation, we can write the modification to the dissipative dynamics of the internal levels of the particle as follows
\eqn{
\Delta\mc{L}_P\sbkt{\hat \rho_P} =& \frac{\Delta \Gamma_e}{2}\left[2\hat{\sigma}_-\hat \rho_P\hat{\sigma}_+-\hat{\sigma}_+\hat{\sigma}_-\hat\rho_P-\hat \rho_P\hat{\sigma}_+\hat{\sigma}_-\right]\non\\
&+\frac{ \Delta\Gamma_m}{2}\left[2\hat{S}_-\hat \rho_P\hat{S}_+ - \hat{S}_+\hat{S}_-\hat \rho_P-\hat \rho_P\hat{S}_+\hat{S}_-\right],
}
where the corrections to the dissipation rates for the  excited states are given as
\eqn{\label{dgammae}
\Delta\Gamma_e& = \frac{4\mu_0\omega_e^2\abs{\mb{d}}^2 }{\hbar} \im\sbkt {\bkt{{\bf e}_+}^T {\dbar{G}_{\mr{sc}}\bkt{{\bf r}_0, {\bf r}_0, \omega_e}}\,{\bf e}_-},\\
\label{dgammam}
\Delta\Gamma_m & =-\frac{2{\hbar}\mu_0 \gamma_0^2}{\hbar} \, \im\sbkt{\bkt{{\bf e}_+}^T {\dbar{\mc{G}}_{\mr{sc}}\bkt{{\bf r}_0, {\bf r}_0, \omega_m}}\,{\bf e}_-}.
}
Here $\Delta\Gamma_e $ refers to the correction to the dissipation for the excited state $\vert{b}\rangle$, and $\Delta\Gamma_m( \langle{S, m_S}\vert\hat{S}^+\hat{S}^-\vert S, m_S\rangle) $  is the modified spin flip rate for transition between  the magnetic sublevel $\vert{S, m_S}\rangle\rightarrow\vert{S, m_S-1}\rangle$ in the presence of the surface \cite{spinflip1, spinflip2, superchip}.

\section{Casimir-Polder force on a ground state magnetic particle}
\label{CPground}

Having  obtained the general expressions for the surface-induced dispersive Eq.\,\eqref{de-}--Eq.\,\eqref{dmz} and dissipative Eq.\,\eqref{dgammae}--Eq.\,\eqref{dgammam} corrections to the particle dynamics, we now consider  a particle in the ground state $\ket{a, S, m_S = -S}$   placed near a planar surface, and study the resulting vacuum-induced forces.

The level shift for the ground state  can be written as
\eqn{\label{grshift}
\bra{a, S,  -S}\Delta\hat H_P \ket{a, S, -S} =    U_e ^{(-)} +   U_m ^{(-)}+  U_m ^{(z)},
}where we have defined  $ U _e ^{(-)} \equiv   u_e ^{(-)}$, $  U _m ^{(-)} \equiv S   u_m ^{(-)} $, $  U _m ^{(z)} \equiv S^2   u_m ^{(z)} $. Using the Green's tensor for a point dipole near a planar surface, as defined in Eq.\,\eqref{green} and Eq.\,\eqref{greencc},  one obtains
\begin{widetext}
\eqn{\label{degr}
  U_e^{(-)}=&\frac{3\hbar\Gamma_0}{8\pi }\int_0^\infty\frac{\dd\xi}{\omega_e}\, \frac{\xi^2}{\xi^2+\omega_e^2}
\int_{\xi/c}^\infty{\frac{\dd \kappa_\perp}{k_e}}e^{-2\kappa_\perp z_0}\sbkt{r_s\bkt{\kappa_\perp, i\xi} -r_p\bkt{\kappa_\perp, i\xi}\frac{\kappa_\perp^2c^2}{\xi^2} },\\
\label{dmgr}
  U_{m}^{(-)}=& \frac{3\hbar\Gamma_0 }{8\pi}\tilde\omega \eta S \int_0^\infty \frac{\dd\xi}{\omega_e}\frac{\xi^2 }{\xi^2 +\omega_m^2}\int_{\xi/c}^\infty \frac{\dd \kappa_\perp}{k_e}e^{-2\kappa_\perp z_0} \sbkt{r_p \bkt{\kappa_\perp , i\xi }  - r_s \bkt{\kappa_\perp , i\xi } \frac{\kappa_\perp^2c^2 }{\xi^2}},\\
\label{dmgrz}
  U_m^{(z)}=&- \frac{3\hbar\Gamma_0 }{8 }\eta S^2 \lim_{\xi\rightarrow0}\int_{\xi/c}^\infty \frac{\dd\kappa_\perp }{ k_e} e^{-2 \kappa_\perp z_0 } \frac{1}{k_e^2}\bkt{\kappa_\perp^2 +\frac{ \xi^2}{c^2}}r_s\bkt{\kappa_\perp , i\xi }.
}
\end{widetext}
  Here $\Gamma_0\equiv \abs{\mb{d}}^2k_e^3/(3\pi \epsilon_0 \hbar )$ refers to the spontaneous emission rate for the ED transition in free space, $\tilde\omega\equiv \omega_m/\omega_e$, $k_{e}\equiv \omega_{e}/c$, and $r_{s,p}$ are the Fresnel coefficients (see Eq.\,\eqref{fresnel}) for the field reflecting off the planar half space  that include the surface material properties. 
We define the characteristic magnetizability to polarizabilty parameter as \eqn{\label{etaj}
\eta\equiv& \frac{{\hbar}^2\gamma_0^2}{\abs{\mb{d}}^2 c^2}= \frac{ \alpha^2}{{\abs{\mb{d}}^2/\bkt{ea_0}^2}},
}
with $\alpha$ as the fine-structure constant, $e$ as the electronic charge and $a_0$ as the Bohr radius.  The parameter $\eta $ determines the characteristic strength of the magnetic to electric CP potential,  thus playing a key role in deciding the overall sign of the CP force. It can be seen from Eq.\,\eqref{dmgr} and Eq.\,\eqref{dmgrz} that the effective strength of the total magnetic  potential has a factor of $S$ for the off-resonant interaction term, and $S^2$ for the diagonal  $\hat{S}_z $ term in the interaction Hamiltonian. The latter contributes only for surfaces with a non-vanishing static frequency response \cite{Barton}, and the scaling of the potential as $\sim S^2 $ can be understood in terms of the magnetostatic  interaction of the dipole with its  image.

By rewriting the electric and magnetic dipole moments for the particle in terms of a corresponding polarizability, the foregoing analysis is generally applicable to a magnetic particle that can be described as a point electric and magnetic  dipole near a planar surface, such as a  nanosphere with a radius smaller than other relevant length scales in the problem \cite{nanoCP}. The electric and magnetic polarizability tensors for the magnetic particle in its ground state can be written as follows \cite{Buh1}
\eqn{\label{alphae}
\dbar{\alpha}_e \bkt{\omega} &= \frac{\abs{\mb{d}}^2\omega_e}{\hbar\bkt{\omega_e^2-\omega^2} }\mb{e}_+\mb{e}_- ,\\
\label{alpham}
\dbar\alpha_m \bkt{\omega} & = \frac{2\hbar \omega_m \gamma_0^2 S}{\omega_m^2-\omega^2 }\mb{e}_+\mb{e}_-  + \pi\hbar \gamma_0 ^2 S^2 \delta(\omega) \mb{e}_z\mb{e}_z.
}
We note that the static electric polarizability can be related to the free space spontaneous emission as $\alpha_e(0) = 3\pi\epsilon_0 \Gamma_0 /\bkt{k_e^4 c}$.
\subsection{Ground state particle near a perfect conductor}
\label{groundPC}
For the case of a particle placed near a perfect conductor, such that the reflection coefficients are  $r_p = 1$ and $r_s = -1$ \cite{Buh1}, the level  shifts Eq.\,\eqref{degr}, Eq.\,\eqref{dmgr} and Eq.\,\eqref{dmgrz} can be written as

\eqn{\label{degrpc} \tilde U _e^{(-)}&=-\frac{3}{32\pi\tilde{z}^3 }\int_0^\infty\frac{\dd\xi\, \omega_e}{\xi^2+\omega_e^2}f\bkt{\frac{2\xi z_0 }{c}}
,\\
\label{dmgrpc}
  \tilde U_{m}^{(-)}&=\frac{3}{32\pi\tilde{z}^3} {\eta S}{} \int_0^\infty \frac{\dd\xi\,\omega_m}{\xi^2 +\omega_m^2} f\bkt{\frac{2\xi z_0}{c}},\\
\label{dmgrpcz}
  \tilde U _m^{(z)} &=\frac{3 }{32\tilde{z}^3} \eta S^2.
}
We have defined the dimensionless particle-surface distance $\tilde z\equiv k_e z_0$,  the dimensionless  potentials $   \tilde U_{e,m}^{(\pm, z)} \equiv   U_{e,m}^{(\pm, z)}/(\hbar \Gamma_0 )$, and $f(x)\equiv ({1+x+x^2})e^{-x}$. As an important point, we note that while the electric CP potential is attractive, the magnetic CP potential is repulsive. This can be understood from the method of images in electromagnetism, since  an electric dipole placed near an infinite planar perfect conductor is attracted towards its image, while a magnetic dipole experiences a repulsive force \cite{LennardJones, CP1948}. We further note that the static contribution to the magnetic level shift Eq.\,\eqref{dmgrpcz} coincides with the energy of a static magnetic dipole interacting with its  image near a perfectly conducting planar surface  \cite{saslow}. 

We consider the  CP potentials $U_{e,m}^{(-)}$ in two asymptotic limits depending on the particle-surface distance relative to the characteristic  length scales for ED and MD interactions. For example,
in the  non-retarded  limit where $z_0 \ll c/\omega_{e (m)}$, the corresponding electric (magnetic) CP potential interaction scales as $\sim 1/z_0^3$. In the
retarded limit, given by $z_0\gg c/\omega_{e (m)} $, the electric (magnetic)  CP potential scales as $\sim 1/z_0^4$. Further assuming that $\omega_e\gg\omega_m$ \footnote{We restrict our attention here to magnetic dipole transitions within a single electronic level, which typically occur at microwave or radio frequencies, unlike optical magnetic dipole transitions \cite{Eu3Novotny} that can occur between different electronic levels.}, one can define the three regimes depicted in Table\,\ref{table1} as follows.
\begin{itemize}
\item {Region I, defined by $z_0\ll c/\omega_{e,m}$, wherein both the electric and magnetic CP forces are non-retarded (NR), and defined as
\eqn{\label{lim1}
\tilde{U}_{e}^{(\mr{NR})}&\equiv \lim_{z_0 \ll c/\omega_e }\tilde {U}_e^{(-)},\non\\
\tilde{U}_{m}^{(\mr{NR})}&\equiv \tilde {U}_m ^{(z)}+\lim_{z_0 \ll c/\omega_m }\tilde {U}_m^{(-)}.
}
}
\item {Region III, defined by $z_0\gg c/\omega_{e,m}$ such that both the electric and magnetic CP potentials are retarded (R), such that
\eqn{\label{lim3}
\tilde{U}_{e}^{(\mr{R})}&\equiv \lim_{z_0 \gg c/\omega_e }\tilde {U}_e^{(-)},\non\\
\tilde{U}_{m}^{(\mr{R})}&\equiv \tilde {U}_m ^{(z)}+\lim_{z_0 \gg c/\omega_m }\tilde {U}_m^{(-)}.
}
}
\item {Region II, defined by  $c/\omega_{e}\ll z_0\ll c/\omega_{m}$ such that the electric CP potential is retarded (R), while the magnetic CP potential is non-retarded (NR).
}

\end{itemize}
The CP potentials in these asymptotic limits are given in Table\,\ref{table1}.

Let us define the dimensionless  total force on the particle as \eqn{\label{ftot}
\tilde{\mc{F}}_{\mr{tot}} = \tilde{\mc{F}}_V+ \tilde{\mc{F}}_{\mr{G}},} where $\tilde {\mc{F}}_V$ refers to the vacuum-induced   force and $\tilde{\mc{F}}_\mr{G}$ to the gravitational force, assumed along the negative $z$-direction. Assuming the  mass of the particle to be $M = S m_u $, where $m_u \approx 1.66\times 10^{-27 }$\,kg refers to the atomic mass unit, the dimensionless gravitational force is defined as $\tilde{\mc{F}}_{\mr{G}} \equiv  -M g/(\hbar\Gamma_0k_e)$, $g$ being the acceleration due to gravity.

The dimensionless vacuum-induced force due to the ED and MD interactions is given by $\tilde{\mc{F}}_V\equiv  \tilde{\mc{F}}_e^{(-)}+\tilde{\mc{F}}_m^{(-)}+\tilde{\mc{F}}_m^{(z)},$
where $  \tilde{\mc{F}}_{e,m}^{(j)}\equiv -\partial \tilde U_{e,m}^{(j)}/\partial\tilde z $. We further define the total force excluding the magnetostatic contribution as the sum of  the CP forces and gravity
\eqn{\label{ftot-}
\tilde{\mc{F}}^\mr{CP}_{\mr{tot}}\equiv \tilde{\mc{F}}_e^{(-)}+\tilde{\mc{F}}_m^{(-)}+\tilde{\mc{F}}_\mr{G},}
which is relevant for surfaces with a vanishing static frequency response.

 In the following, we analyze the total force with and without the magnetostatic contribution as given by Eq.\,\eqref{ftot} and Eq.\,\eqref{ftot-} respectively.
For estimate purposes, we choose $\omega_e = 2\pi\times 10^{15}$\,
Hz, $\omega_m = 2\pi\times 10^{10}$\,Hz, and $\Gamma_0 = 18$\,MHz, which corresponds to an ED moment value of $\abs{\mb{d}}\approx ea_0/2$.

\begin{figure*}
\begin{center}
\subfloat[]{\includegraphics[width = 3.5 in]{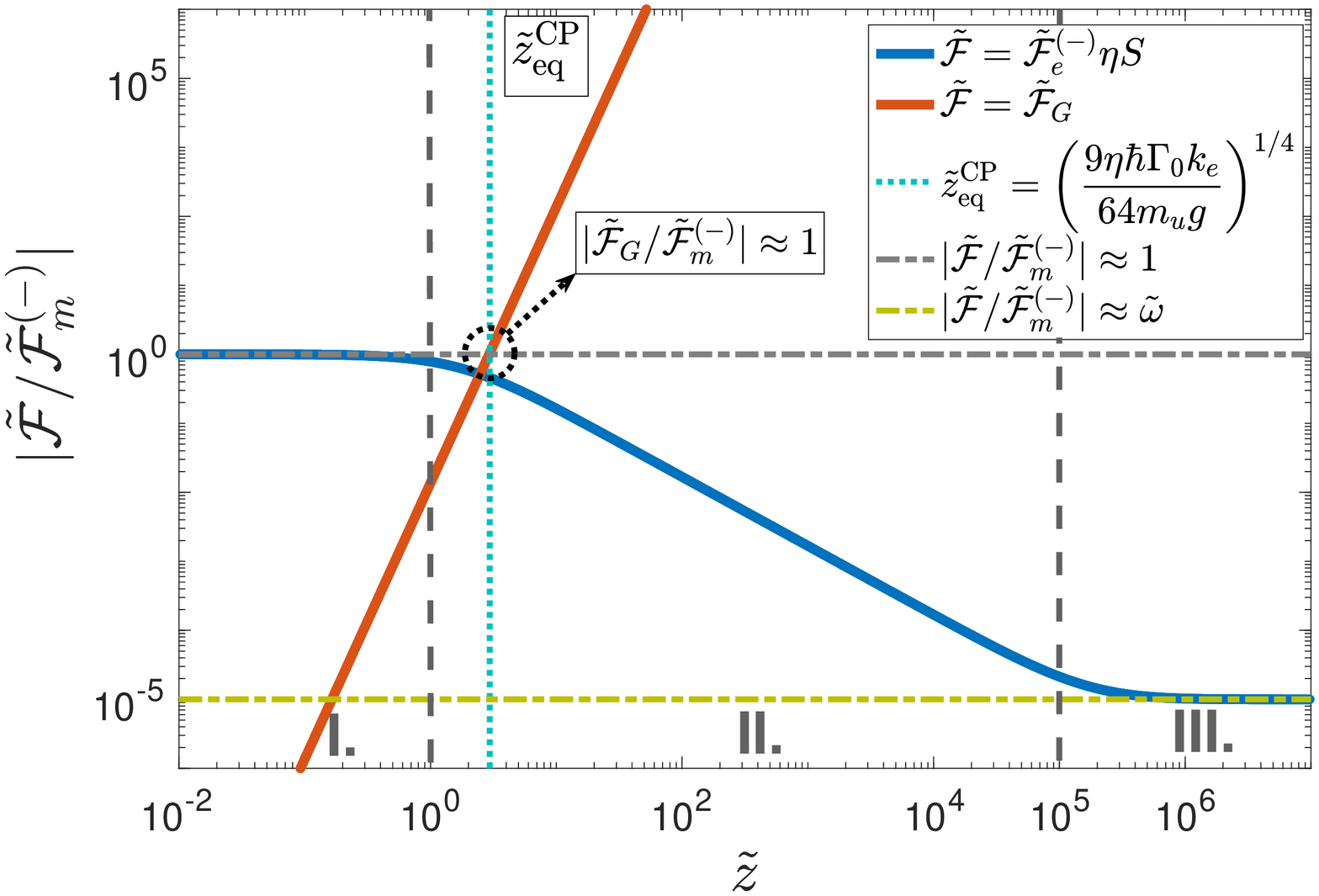}}
\subfloat[]{\includegraphics[width = 3.5 in]{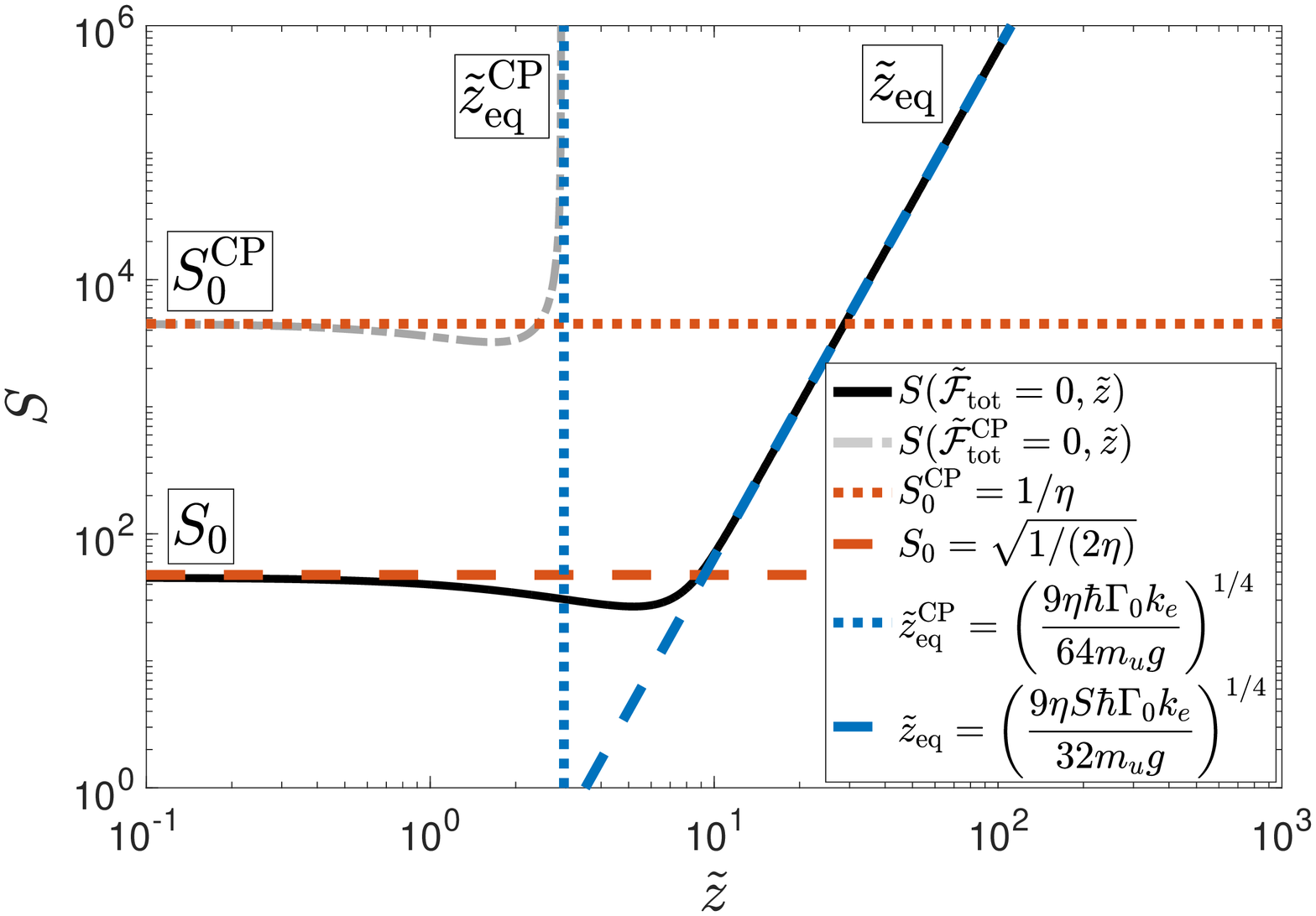}}
\end{center}

\caption{ (a) Relative magnitudes of the rescaled electric CP force (blue solid) and gravity (red solid) compared to the magnetic CP force on a ground state particle as a function of distance. The three distance regimes as shown in Table\,\ref{table1} are depicted as the dashed gray vertical lines. In region I (III), the electric and magnetic CP forces are such that $\tilde{\mc{F}}_m^{(-)}/\tilde{\mc{F}}_e^{(-)}\approx\eta S$ $(\tilde{\mc{F}}_m^{(-)}/\tilde{\mc{F}}_e^{(-)}\approx \eta S/\tilde\omega)$, yielding the condition given by Eq.\,\eqref{constCP} (Eq.\,\eqref{constCPfar}) for near-(far-)field Casimir repulsion. For a large enough spin, the condition $|\tilde{\mc{F}}_G/\tilde{\mc{F}}_m^{(-)}|\approx1$ defines the condition for achieving a stable equilibrium, as shown by the intersection point for the red solid and gray dashed-dotted curves. It can be seen that this point coincides with the approximate analytical expression obtained in Eq.\,\eqref{z2cp} as depicted by the vertical blue dotted line.  (b) Spin value $S(\tilde{\mc{F}}_{\mr{tot}}^{(\mr{CP})}=0,\tilde z)$ as a function of the particle-surface distance $\tilde z$ for the total force $\tilde{\mc{F}}_{\mr{tot}}^{(\mr{CP})}$ with(out) the magnetostatic contribution to be zero shown in black solid (gray dashed-dotted) curve. The threshold spin value $S_0^{(\mr{CP})}$ for near field repulsion defined by (Eq.\eqref{constCP}) Eq.\eqref{const} is shown by the red (dotted) dashed line. The approximate stable equilibirum point  given by  (Eq.\,\eqref{z2cp}) Eq.\,\eqref{z2} are shown in blue (dotted) dashed curve.  }
\label{FigPC}
\end{figure*}

\begin{itemize}
\item{Casimir-Polder repulsion (excluding magnetostatic contribution): 
 The threshold value of the spin to facilitate an overall repulsive CP force, given by Eq.\,\eqref{ftot-}, in the region I  is roughly given as
\eqn{\label{constCP}\eta S_0^\mr{CP}\approx1\text{, or } S_0^\mr{CP}\approx10^4,}
as can be seen from  Fig.\,\ref{FigPC}\,(a). Thus, the condition $S\gtrsim S_0^\mr{CP}$ defines a key constraint for achieving near-field CP repulsion  for a magnetic particle in the ground state.
In the retarded regime (region III), the condition for having a repulsive CP force becomes \eqn{\label{constCPfar}\eta S\gtrsim \tilde\omega,} as seen from Fig.\,\ref{FigPC}\,(a). This can be readily realized for $\tilde\omega\lesssim10^{-4}$, thus yielding a repulsive force in the far-field limit.

For $S>S_0^\mr{CP}$, approximating the total force as $\tilde{\mc{F}}_{\mr{tot}}^\mr{CP}\approx\tilde{\mc{F}}_m^{(-)} +\tilde{\mc{F}}_\mr{G}$, the position for having a stable equilibrium is given as
\eqn{\label{z2cp}
\tilde z_\mathrm{eq}^\mr{CP} =\left(\frac{9\eta\hbar\Gamma_0 k_e}{64  m_ug}\right)^{1/4}.
}
We note that with both the repulsive magnetic CP $(\tilde{\mc{F}}_m^{(-)})$ and attractive gravitational $(\tilde{\mc{F}}_\mr{G})$  contributions to the total force scaling as $\sim S$, the equilibrium position is independent of the spin, as seen from the dotted vertical line in Fig.\,\ref{FigPC}\,(b). It can be also verified from Fig.\,\ref{FigPC}\,(a), that the gravitational force becomes comparable in magnitude to the magnetic CP force for $\tilde z\approx\tilde z_{\mr{eq}}^{\mr{CP}}$. }

\item{Total vacuum-induced repulsion (including magnetostatic contribution): Considering the total vacuum-induced force along with gravity, as defined in \eqref{ftot}, we see from Table\,\ref{table1} that   to achieve a repulsive near-field total force one requires $\eta S\bkt{2S+1}\gtrsim1$. Assuming that $\vert\mb{d}\vert \approx e a_0/2$, we obtain a threshold spin value of \eqn{\label{const}S_0\approx\sqrt{1/(2\eta)}\text{, or } S_0\approx50,} for near field repulsion. 

 For ${S}>S_0 $, the total force can be  well-approximated by $\tilde{\mc{F}}_{\mr{tot}}\approx\tilde{\mc{F}}_{{m}}^{(z)} +\tilde{\mc{F}}_{{G}}$, such that  with the near-field magnetostatic repulsion and far-field gravitational attraction, one obtains a stable equilibrium point  \eqn{\label{z2}\tilde z_{\text{eq}}\approx \bkt{\frac{9 \eta S \hbar\Gamma_0 k_e}{32 m_u g} }^{1/4}.} 
We note from the above that the equilibrium point gets pushed away from the surface on increasing the spin value $S$.}
\end{itemize}

\begin{table*}[htb]
\begin{center}
\begin{tabular}{|p { 3 cm}|p { 4.5 cm}|p { 4.5 cm}|p { 4.5 cm}|}
\hline
      &{\begin{center}
     Perfect conductor 
      \end{center} }&{\begin{center}
Drude model
      \end{center}}&{\begin{center}
Plasma model
      \end{center}} \\
\hline
     \begin{center}
     I.
     \begin{align*} {z_0 \ll c/\omega_{e,p,m}}
     \end{align*}
     \end{center}&
     \begin{center}$\begin{aligned}
       \tilde U^{\mr{NR}}_e &\approx-\frac{3}{64 \tilde{z}^3}\\
  \tilde U^{\mr{NR}}_m &\approx\frac{3}{64 \tilde{z}^3}\eta S\bkt{2S+1}\end{aligned}$\end{center}
     & \begin{center}{$\begin{aligned}
 \tilde U^{\mr{NR}}_{e} &\approx\frac{\mc{C}_{e,D}^{(3)}}{\tilde{z}^3}
\\   \tilde U_{m}^{\mr{NR}} &\approx\frac{\mc{C}_{m, D}^{(1)}}{\tilde z}\end{aligned}$}\end{center}&\begin{center}$\begin{aligned}
  \tilde U^{\mr{NR}}_{e} &\approx\frac{\mc{C}_{e,D}^{(3)}}{\tilde{z}^3}
\\   \tilde U_{m}^{\mr{NR}} &\approx\frac{\mc{C}_{m, P}^{(1)}}{\tilde z}  \end{aligned}$\end{center}\\
\hline
     \begin{center}
     II.          \begin{align*}
         { c/\omega_{e,p}\ll z_0\ll c/\omega_m}
     \end{align*}
     \end{center}&\begin{center}$\begin{aligned} \tilde U^{\mr{R}}_e& \approx-\frac{3}{16\pi \tilde{z}^4}\\
      \tilde U^{\mr{NR}}_m &\approx\frac{3}{64 \tilde{z}^3}\eta S \bkt{2S+1}\end{aligned}$  \end{center}& 
\begin{center}    $ \begin{aligned} \tilde U^{\mr{R}}_e& \approx-\frac{3}{16\pi \tilde{z}^4}\\
      \tilde U^{\mr{Int}}_m &\approx\frac{3}{64 \tilde{z}^3}\eta S  \end{aligned}$\end{center}&
     \begin{center}$
\begin{aligned}
 \tilde U^{\mr{R}}_e &\approx-\frac{3}{16\pi \tilde{z}^4}\\
 \tilde U^{\mr{Int}}_m &\approx\frac{3}{64 \tilde{z}^3}\eta S \bkt{2S+1}
\end{aligned}$
\end{center}
\\
\hline
\begin{center}
III.
\begin{align*}
         { c/\omega_{e,p,m}\gg z_0}
     \end{align*}
\end{center}     &\begin{center}
$\begin{aligned}
 \tilde U^{\mr{R}}_e &\approx-\frac{3}{16\pi \tilde{z}^4}\\
 \tilde U^{\mr{R}}_m &\approx\frac{3}{16\pi \tilde{z}^4}\frac{\eta{S}}{\tilde{\omega}}\bkt{\frac{\pi S\tilde{z}\tilde{\omega}}{2}+1}
\end{aligned}$
\end{center}&\begin{center}
$\begin{aligned}
 \tilde U^{\mr{R}}_e &\approx-\frac{3}{16\pi \tilde{z}^4}\\
 \tilde U^{\mr{R}}_m &\approx\frac{3}{16\pi \tilde{z}^4}\frac{\eta{S}}{\tilde{\omega}}
\end{aligned}$
\end{center} &\begin{center}
$\begin{aligned}
 \tilde U^{\mr{R}}_e &\approx-\frac{3}{16\pi \tilde{z}^4}\\
 \tilde U^{\mr{R}}_m &\approx\frac{3}{16\pi \tilde{z}^4}\frac{\eta{S}}{\tilde{\omega}}\bkt{\frac{\pi S\tilde{z}\tilde{\omega}}{2}+1}
\end{aligned}$
\end{center}\\
\hline
\end{tabular}
\caption{Energy level shifts  for a  magnetic particle in the ground  state $\ket{a, S, m_S = -S}$ placed near a (1) perfectly conducting planar surface, and a metal surface described by the (2) Drude, and (3) plasma models for dielectric permittivity. The level shifts  in (non-)retarded asymptotic limits are defined as in Eq.\,\eqref{lim1} and Eq.\,\eqref{lim3}. For the Drude and plasma models, there being a separate length scale corresponding to the plasma frequency of the surface, there is a different asymptotic behavior for the magnetic CP potential in the intermediate distance regime where one is in the non-retarded limit relative to the MD transition frequency but retarded relative to the plasma frequency of the surface $(c/\omega_{ p}\ll z_0\ll c/\omega_m )$, as defined in Eq.\,\eqref{limint}.  It can be seen that in the retarded limit, the level shifts for the plasma (Drude) model mimic those for the perfect conductor, with (without) the static frequency contribution, since in the far-field limit the response of the metal surface at low frequencies corresponds to that of a perfect conductor. The coefficients $\mc{C}_j^k$s for Drude and plasma models are defined in Eq.\,\eqref{ce3D}--Eq.\,\eqref{cm1P}.}
\label{table1}
\end{center}
\end{table*}

We have discussed here the general conditions and parameter regimes for realizing a repulsive  force between a ground state magnetic particle and a perfectly conducting planar surface using MD interaction and the possibility of levitating it against gravity. In the following section we study how these conditions generalize for the particle placed near more realistic metal surface described by the Drude or plasma model.

\subsection{ Ground state particle near metal surfaces}
\label{differentsurfaces}
The dielectric permittivity of a metal  can be modeled by the Drude or  plasma models as described by the permittivity functions as in Eq.\,\eqref{drude} and Eq.\,\eqref{plasma} respectively. The key difference in the two models is that the plasma model disregards the relaxation of conducting electrons in the metal. This difference matters the most for the low frequency response of the metals, and can lead to very different resulting Casimir forces \cite{Hartmann, QVForcesBook, Mostepanenko}. Such a discrepancy has been much debated in the literature, with the experimental results  favoring the plasma model in some cases, and Drude in others \cite{Decca1,Decca2,  Mohideen1,Mohideen2, Mohideen3, Sushkov, Tang}.

Particularly, the response of the metal at zero frequency is largely different between the two models. For example, while the Fresnel coefficient $r_s(\kappa_\perp, 0)\rightarrow0$ for the Drude model, it is non-vanishing for the plasma model \cite{Mostepanenko}. As we have seen from  Eq.\,\eqref{dmgrz}, the Fresnel coefficient $r_s(\kappa_\perp, 0)$ is crucial for determining the magnetostatic contribution. Thus,  for a ground state particle, given that  $\hat S_z$ term contribution scales as $\sim S^2$, as opposed to the broadband CP contribution  that scales as $\sim S$ (see Eq.\,\eqref{grshift}), the two models for the dielectric permittivity of the metal could yield a very different  magnetic  force  for a large enough spin system. In the following sections we study the conditions for achieving an overall repulsive  force from the ED and MD interactions near a metal surface described by the Drude and plasma model, particularly noting the difference in the two results due to the magnetostatic contribution.
\vspace{0.5 cm}

\subsubsection{Drude model}

Considering a metal surface with a Drude model for dielectric permittivity 
\eqn{\label{drude}
\epsilon_D\bkt{i\xi} = 1+\frac{\omega_p^2}{\xi^2+\gamma\xi},
}
where $\omega_p$ is the plasma frequency of the metal and $\gamma\ll\omega_p $ corresponds to the loss. Using the above dielectric permittivity for the optical response of the surface, one can find the ground state level shifts as given by Eq.\,\eqref{degr} and Eq.\,\eqref{dmgr}. Assuming $\omega_{p,e}\gg\omega_m$ \cite{Skagerstam}, there are three limiting cases of the resulting level shifts as summarized in Table\,\ref{table1}. In region (I) III, both the ED and MD interaction induced level shifts are in the (non-)retarded limit.  In addition to the retarded and non-retarded regimes, there is a separate length scale in the problem that corresponds to the plasma frequency of the metal.  This leads to a different asymptotic behaviour for the magnetic level shift in the intermediate regime demarcated by $c/\omega_p\ll z_0\ll c/\omega_m$, defined as 
\eqn{\label{limint}\tilde{U}_m^{\mr{Int}}\equiv \tilde {U}_m^{(z)}+\lim_{c/\omega_p\ll z_0\ll c/\omega_m }\tilde{U}_m^{(-)}.} We note that in the retarded limit the metal surface behaves as a perfect conductor, as far as the electric level-shift is concerned. This can be physically understood by considering that  the relevant frequencies for the far-field interaction being small, the plasma frequency of the metal appears infinite, which corresponds to the response of a perfect conductor.

Importantly, it can be seen from Eq.\,\eqref{fresnel} that with $\lim_{\xi\rightarrow0}r_s(\kappa_\perp,i\xi) = 0$, the static  contribution as given by Eq.\,\eqref{dmgrz} vanishes, such that the total magnetic CP potential is given as
\begin{widetext}
\eqn{
  \tilde U_{m}^{(-)}=& \frac{3 }{8\pi}\tilde\omega \eta S \int_0^\infty \frac{\dd\xi}{\omega_e}\frac{\xi^2 }{\xi^2 +\omega_m^2}\int_{\xi/c}^\infty \frac{\dd \kappa_\perp}{k_e}e^{-2\kappa_\perp z_0} \sbkt{r_p \bkt{\kappa_\perp , i\xi }  - r_s \bkt{\kappa_\perp , i\xi } \frac{\kappa_\perp^2c^2 }{\xi^2}},
}
\end{widetext}
which can be calculated in the three asymptotic limits using the approach in Appendix\,\ref{appB}, as summarized in Table\,\ref{table1}.   Specifically, we note here that the above potential scales linearly with the total spin $S$, in contrast to the case of a perfect conductor.

It can be seen from Table\,\ref{table1}, that for the total CP force to be repulsive in region I, one requires \eqn{\label{constD}\abs{\frac{{ }\tilde U_m^{\mr{NR}}} {{ }\tilde U_e^{\mr{NR}}}}& \approx \frac{\mc{C}_{m,D}^{(1)}}{\mc{C}_{e,D}^{(3)}}\tilde{z}^2\sim \eta S \frac{\omega_p^2}{\omega_e^2}\frac{\ln\bkt{\gamma/\omega_m}}{\gamma/\omega_m}\tilde z^2,}
with the coefficients $\mc{C}_{m,D}^{(1)}$ and $\mc{C}_{e,D}^{(3)}$ given by  Eq.\,\eqref{cm1D} and  Eq.\,\eqref{ce3D}, assuming 
 $\omega_p\sim100 \gamma\sim 10\omega_e \sim10^7\omega_m$. Considering the particle near a gold surface with $\omega_p\approx1.36\times 10^{16}$\,Hz and $\gamma\approx10^{14}$\,Hz, one requires \eqn{\label{SD}\eta {S}\gtrsim \frac{10^3}{\tilde z^2},} for the total CP force in region I to be repulsive. For a particle-surface separation $z_0\approx10$\,nm, we find that one requires the total spin to be as large as $S\gtrsim 10^8$ to achieve repulsion.

\subsubsection{Plasma model}
We now consider the plasma model for dielectric permittivity, by setting the relaxation parameter in the Drude model to zero
\eqn{\label{plasma}
\epsilon_P\bkt{i\xi} = 1+\frac{\omega_p^2}{\xi^2}.
}
Using the above dielectric response in Eq.\,\eqref{degr} and Eq.\,\eqref{dmgr}, one can find the ED and MD interaction induced potentials in the three asymptotic limits as outlined in Table\,\ref{table1}. In the near-field limit, one can thus write the ratio of the repulsive magnetic potential to the attractive electric part as
\eqn{\label{constP}
\abs{\frac{{ }\tilde U_m^{\mr{NR}}} {{ }\tilde U_e^{\mr{NR}}}}& \approx \frac{\mc{C}_{m,P}^{(1)}}{\mc{C}_{e,D}^{(3)}}\tilde{z}^2\sim \eta S \bkt{2S+1} \frac{\omega_p^2}{\omega_e^2}\tilde z^2,
}
where the coefficients $\mc{C}_{m,P}^{(1)}$ and $\mc{C}_{m,D}^{(3)}$ are given by Eq.\,\eqref{cm1P} and Eq.\,\eqref{ce3D} respectively, assuming 
 $\omega_p\sim 10\omega_e \sim10^7\omega_m$. Again, considering the particle near a gold surface with $\omega_p\approx1.36\times 10^{16}$\,Hz, one requires
\eqn{\label{SP}
\eta S\bkt{2S +1} \gtrsim \frac{10^{-2}}{\tilde z^2},
}
which yields, assuming a particle-surface separation of $z_0 = 10$\,nm, a spin value of $S\gtrsim 10^2$ for the total force in the near-field regime to be repulsive. Thus, we see that one requires a much smaller spin value to achieve an overall repulsive force for a particle near a  metal surface if the surface response at zero frequency  is described by the plasma model as opposed to the Drude model. Further, comparing Eq.\,\eqref{constD} and Eq.\,\eqref{constP}, we see that the repulsive magnetic level shift is for the Drude model relative to the plasma model is smaller by a factor $\sim \omega_m/(S\gamma).$


Considering that one requires spin values as large as $S\sim 10^8$ to achieve an overall repulsive CP force near a metal surface described by the Drude model, we further propose some  possible ways to boost the repulsive magnetic contribution to the total CP force in the following section.

\section{Casimir-Polder force on an excited magnetic particle}
\label{CPexcited}

As discussed in section \ref{model}, for a particle in the ground state, its only possible  interactions with the field are via virtual excitation processes that can occur at all frequencies, making the interaction essentially broadband. However, for an particle in the excited state, there can be a real transition to lower energy states accompanied by the emission of a real photon, which yields a resonant shift to the excited state and modifies the dissipation rate in the presence of a surface. With the resonant shift depending singularly on the response of the surface at the transition  frequency of the particle as opposed to a broadband  response, as a result the excited state shifts can be manipulated relatively easily by altering the density of field modes at the transition frequency \cite{KSDEC}.

In addition, we further note from Eq.\,\eqref{dHa} (and Eq.\,\eqref{Spm}) that  if we consider the magnetic particle to be in the excited level $\ket{S, m_S =0} $, the characteristic strength for the magnetic CP interaction (excluding the static contribution) is larger by a factor of $S$ in comparison to that for the ground state. Drawing the analogy between a spin system and the  Dicke model \cite{Dicke, Haroche}, such an enhancement can be understood as that in the case of  superradiance  in a collection of atoms. In the case of  Dicke superradiance, the enhancement in the transition amplitude for the superradiant state corresponding to $\ket{S, m_S = 0}$ leads to a boost by a factor of $S$ in the collective spontaneous emission rate of the atoms. Since the CP force is also a fluctuation phenomenon, one can naturally expect that the cooperative effects that influence the spontaneous emission for a collection of atoms to also influence the CP  force \cite{CollectiveCP, Vcollective}.

Thus, given that (i) the excited state shifts depend  on the response of the surface at the resonant transition frequency, such that one can enhance the repulsive magnetic CP potential by appropriately engineering the surface response, and (ii) that by preparing the particle in an excited state one can further boost the repulsive magnetic CP force, we therefore consider the CP interaction for a particle in the state $\ket{a, S, m_S = 0}$.  We analyze these two effects in the following.

\subsection{The magnetic CP potential for the excited sublevel $\vert{S, m_S = 0}\rangle$}

The total level shift of the state $\ket{S, m_S = 0}$ comprises of three contributions. First, there is an off-resonant contribution from the virtual transition to the level $\ket{S, m_S = +1}$ as described by Eq.\,\eqref{dm-}. Next, considering   the transition to the level $\ket{S, m_S = -1}$ there is an off-resonant and a resonant contribution as seen from the first and second terms in Eq.\,\eqref{dm+} respectively.

Adding together the contributions from the transition to the levels $\ket{S,m_S = \pm1}$ using Eq.\,\eqref{dm+} and Eq.\,\eqref{dm-}, we find the total magnetic CP shift on the level $\ket{S, m_S = 0}$  for a particle near a planar surface ,
\eqn{\label{Dm0}
{ }\tilde U_{m}^{(0)}= -\frac{3\eta S\bkt{S+1}\tilde {\omega}^3}{16}\re \int_0^\infty\frac{\dd k_{\parallel}}{k_m}\,\frac{k_{\parallel} }{\kappa_\perp}e^{-2\kappa_\perp z_0}& \non\\
\sbkt{r_p (\kappa_\perp, \omega_m)  +r_s(\kappa_\perp, \omega_m)\frac{\kappa_\perp^2}{k_m^2}}.&
}
 Comparing the above with  the broadband contribution in Eq.\,\eqref{dmgr}, the magnetizability to polarizability for the $\ket{S, m_S = 0}$ level is larger by an additional factor of $S$ with respect to the ground state. We also note that for $m_S =0$, the contribution from the diagonal interaction term in Eq.\,\eqref{dmgrz} vanishes. 

Considering the particle placed near a perfectly conducting surface, we write the  magnetic CP potential for the state $\ket{a, S, m_S =0}$ as
\eqn{
  \tilde U_m^{(0)} &= \frac{3\eta S\bkt{S+1}}{64\tilde{z}^3}\non\\& \sbkt{\cos  (2 \tilde\omega \tilde{z})+{2\tilde{\omega}\tilde z }\sin  (2 \tilde{\omega}  \tilde{z})-{4\tilde{\omega}^2 \tilde z^2}\cos  \bkt{2 \tilde{\omega}  \tilde{z}}}.
}
In the  non-retarded regime for the magnetic CP potential  $\bkt{\tilde\omega\tilde{z}\ll1}$, the above potential reduces to $ \tilde U_m^{(0),\text{NR}}\approx{ 3\eta S(S+1)}/\bkt{64\tilde{z}^3}$. Thus for the total CP force to be repulsive in the near-field regime such  that $ \tilde U_e^{\mr{NR}}+ \tilde U_m^{(0),\text{NR}}>0$, one requires $\eta S(S+1) \gtrsim1$, or  $S\gtrsim10^2$.


Adding together  the CP force and gravity as before, the total potential in the near-field regime (considering $S\gtrsim10^2$) can be described approximately as the sum of the non-retarded magnetic CP potential and the gravitational  potential as 
\eqn{\label{Ftotapp}\tilde{\mc{F}}_{\mr{tot}}^{\mr{approx}}\approx\frac{9\eta S\bkt{S+1}}{64\tilde{z}^4}-\frac{Mg}{\hbar\Gamma_0k_e}.
}
Thus there is a stable equilibrium point along the $z$-direction at
\eqn{\label{z2ex}\tilde{z}^{(0)}_{\mr{eq} }\approx \bkt{\frac{9\eta S\hbar\Gamma_0 k_e}{64  m_ug}}^{1/4},
}where the attractive  gravitational force balances the magnetic repulsion. We see that on increasing $S$, the magnetic repulsion increases more than the gravitational attraction and the equilibrium point gets pushed further away from the surface.


The modified spin-flip rate given by Eq.\,\eqref{dgammam} in the non-retarded limit is given as
$
\Delta\Gamma_m^{(0), \mr{NR}} \approx \Gamma_0 \eta S(S+1)\tilde {\omega}^3/3\sim 10^{-9}$\,Hz  \cite{spinflip1, spinflip2, superchip}, considering a spin value of $S = 100$. In the absence of surface losses, the decay rate is independent of the particle-surface separation in the non-retarded limit, as
expected from image theory \cite{WylieSipe1}.

In the following section we study the possibility of combining such a ``super-radiant'' enhancement to the repulsive CP force together with an additional boost coming from surface resonance for a particle placed near a metal surface described by the Drude model.

\subsection{Enhancing the magnetic CP force using surface resonances}
It can be seen  from Eq.\,\eqref{Dm0} that the off-resonant shift  contribution from the virtual transition to the level $\ket{S, m_S = +1}$ cancels with that to the level $\ket{S, m_S = -1}$, as expected from second-order perturbation theory, such that the overall magnetic shift on the $\ket{S, m_S =0 }$ level contains only the resonant contribution. As a result, the total magnetic CP potential depends singularly on  the response of the surface at the resonant transition frequency $\omega_m$. Thus, we consider here the  possibility of boosting the magnetic CP force resonantly by engineering the response of the surface at $\omega_m$ following the approach in  \cite{KSDEC}.

Modeling the surface response using the Drude model given by Eq.\,\eqref{drude},  one can  see from Eq.\,\eqref{Dm0} and Eq.\,\eqref{fresnel} that in order to find the CP potential one only requires the dielectric response of the surface at the resonance transition frequency $\epsilon_m\equiv\epsilon(\omega_m)$ (assuming $\mu(\omega_m)=1$). In the non-retarded limit for MD interaction $(\tilde{\omega}\tilde z\ll1)$, we use the approach in Appendix\,\ref{appB} to  write an asymptotic expression for the non-retarded magnetic CP potential near a metal surface as 
\eqn{\label{dmcnr}{
  \tilde U_m^{(0), \mr{NR}} \approx \frac{3\eta S \bkt{S+1} \tilde{\omega}^2}{128\tilde z}\re\sbkt{ \frac{\bkt{\epsilon_m - 1}\bkt{\epsilon_m +5}}{\epsilon_m +1}}}.}
The above potential undergoes a resonance for  $\bkt{\epsilon_m+1}\rightarrow0$ near the plasmon resonance  given by $\omega_m\rightarrow \omega_p/\sqrt{2}$. We define the quality factor for the surface material as $Q\equiv \omega_{p}/(\sqrt{2}\gamma)$, and the dimensionless detuning with respect to the plasmon resonance as $\delta_{p} \equiv \bkt{\omega_m - \omega_{p}/\sqrt{2}}/\gamma$. Assuming a high Q-factor for surface resonance and large detuning such that $Q\gg\abs{\delta_p}\gg1$,  one can rewrite the non-retarded magnetic CP potential as \eqn{\label{Dm0nr}  \tilde U^{(0), \mr{NR}}_m\approx-\frac{3\eta S\bkt{S+1}\tilde\omega^2}{256\tilde{z}} \bkt{\frac{Q}{\delta_p}}.}
The ratio of the magnetic to electric  CP potential given by Eq.\,\eqref{ce3D}  in region I  goes roughly as, \eqn{\abs{\frac{  \tilde U_m^{(0), \text{NR}}}{  \tilde U_e^{\text{NR}}}}\sim\eta S \bkt{S+1} \tilde\omega^2\tilde z^2 \bkt{\frac{Q}{\abs{\delta_p}}},
}
such that to achieve near field repulsion we require
\eqn{
\eta S\bkt{S+1} \gtrsim \frac{1}{\tilde{\omega}^2\tilde z^2}\bkt{\frac{\abs{\delta_p}}{Q}}.
}
For a particle-surface distance of $z_0 = 10 $\,nm, assuming $\abs{\delta_p}\sim 10^2$, $Q\sim 10^{12}$, gives a spin value of $S\sim 10^2$, which is significantly smaller than the spin value required for the case of a ground state particle near a Drude surface $(S\sim 10^8)$. We note that this value is only limited by the quality factor of the surface resonance and can in principle be made arbitrarily small.

The spin-flip rate for the state $\ket{a,S,m_S = 0}$ in the non-retarded limit, near the plasmon resonance of the surface is given by Eq.\,\eqref{dgammam}
\eqn{\label{dgamma0nr}\Delta\Gamma_m^{(0), \mr{NR}}&\approx-\frac{3\Gamma_0\eta S \bkt{S+1} \tilde\omega^2}{16\tilde z}\im\sbkt{\frac{\bkt{\epsilon_m - 1}}{\epsilon_m +1}}\non\\
&\approx\frac{3\Gamma_0\eta S\bkt{S+1} \tilde\omega^2}{64\tilde z} \bkt{\frac{Q}{\delta_p^2}}.
}
For the chosen set of values, assuming $S \sim 10^2$, the decoherence rate becomes $\Delta\Gamma_m^{(0), \mr{NR}}\approx 1$\,MHz.  We note here that the dispersive and dissipative   corrections, given by Eq.\,\eqref{Dm0nr} and  Eq.\,\eqref{dgamma0nr}, contain factors of $Q/\abs{\delta_p}$ and $Q/\abs{\delta_p}^2$ respectively, which are characteristic of a generic coupling between a particle and a resonator \cite{Grimm, KSDEC}. One can thus  find a large enough quality factor $Q$ along with a large detuning $\delta_p$, such that the dispersive shifts can be large, without increasing the spin decoherence rate significantly.

\section{Discussion}
\label{discussion}
In this work, we analyze the possibility of realizing a repulsive  force between a magnetic particle and a planar surface interacting via the vacuum EM field. Considering the toy model of a particle with an electric-dipole transition and a large magnetic spin, we find that as a result of the interplay between the relatively long-ranged repulsive magnetic-dipole contribution and the short-ranged attractive electric-dipole part to the total vacuum-induced force, one can possibly achieve a repulsive interaction for (i) large enough magnetizability to polarizability ratio of the magnetic particle, and (ii) in distance regimes where the ED contribution to the total force is retarded and thus weak, while the MD contribution is non-retarded and can potentially overtake the attractive ED part (see Fig.\,\ref{FigPC}). We find that the  level shifts induced due to the MD interaction for a ground state magnetic particle near a perfectly conducting surface contains two contributions, one  that can be attributed to the broadband Casimir-Polder interaction, and a zero frequency contribution that can be understood as coming from the interaction between the static dipole and its image. Thus, we find that as a fundamental constraint for achieving near-field repulsion via MD interaction for a ground state particle near a perfectly conducting surface one requires an  spin $S$ larger than $\sim1/\alpha$ $(\sim 10^2)$, where $\alpha$ refers to the fine-structure constant, as in  Eq.\,\eqref{const}. In the absence of the static frequency  contribution, the minimum spin value required for achieving a repulsive CP force is $\sim 1/\alpha^2 $ $(\sim 10^4)$, as in  Eq.\,\eqref{constCP}. We formulate  similar conditions and estimates for  achieving Casimir repulsion via MD interaction   for the particle placed near surfaces  described by Drude and plasma  models, given by Eq.\,\eqref{constD} and Eq.\,\eqref{constP}. We also propose some possible ways to enhance the repulsive magnetic CP interaction, such as, by preparing the particle in a superradiant state, and using surface resonances, considering particularly the  level $\ket{S, m_S=0}$. Using the analogy between the Dicke model and a spin system, similar to the enhancement in spontaneous emission, one can understand the   enhancement in the magnetic CP force on the magnetic sublevel level $\ket{S, m_S=0}$ in terms of Dicke superradiance \cite{CollectiveCP}.

Our results could be instructive in identifying potential systems, mechanisms, and regimes where one could realize stable levitation via repulsive magnetic-dipole induced forces. For example,  we remark that the desired spin values of $S\sim 100$ are not far from the large spin values realized  for single molecular magnets (SMMs)  \cite{Molecule1,Molecule2}. In the field of molecular magnetism, there has been a significant interest in creating single molecular complexes with  magnetic spins that can be as large as $S= 105/2$ \cite{Molecule1, Molecule2}. For purposes of a rough estimate, assuming  that an SMM with spin $S\approx50$ constitutes roughly $N\approx10^4$ atoms \cite{Molecule1},  such that its electric polarizability can be approximated as $\alpha_e\sim 10^4 \alpha_0$, $\alpha_0 $  being the single particle electric polarizability as considered earlier, and mass $M\approx 10^4$ amu, we find that such a molecule placed near a perfectly conducting surface could possibly be levitated via the magnetic-dipole interaction.

 Other systems with large quantum spins such as single-domain magnetic nanoparticles \cite{CRORI}, can potentially exhibit a macroscopic spin as large as $S\sim10^5$. We remark that such a particle can be used to differentiate between the zero frequency response  of a metal surface that is described by a Drude  or a plasma model (see Eq.\,\eqref{SD}  and Eq.\,\eqref{SP}). As shown in section \ref{differentsurfaces}, the static frequency contribution being absent for the case of a Drude model, a particle with a large spin sees an MD interaction-induced force that is smaller by a factor $\sim \omega_m/(\gamma S)$ than in the case of a surface described by the plasma model.

Further, identifying collective phenomena as a potential tool to tailor vacuum forces, one can speculate using cooperative effects as a means to probe otherwise weak vacuum forces, or suppress attractive Casimir forces by preparing systems in appropriate collective states \cite{CollectiveCP}.

\acknowledgements{ 
We thank Oriol Romero-Isart and Gerald E. Fux for fruitful discussions and contributions. This work was supported by the European Research Council ( ERC-2013-StG 335489 QSuperMag).
}
\appendix
\section{Medium-assisted EM field}
\label{appendixa}
Using the macroscopic QED formalism \cite{Buh1, Buh2}, the Hamiltonian for the vacuum EM field in the presence of the surface can be written as
\eqn{\label{hv}
H_F =\sum_{ \lambda = e,m}\int \dd^3 r \int \dd\omega\,\hbar\omega\, \hat{\mb{f}}^\dagger_\lambda\bkt{\mb{r}, \omega}\cdot\hat{\mb{f}}_\lambda\bkt{\mb{r}, \omega},
}
with $\hat{\mb{f}}^\dagger_\lambda\bkt{\mb{r}, \omega}$ and ${\hat{\mb{f}}_\lambda\bkt{\mb{r}, \omega}}$ as the bosonic creation and annihilation operators respectively that take into account the presence of the media. Physically these can be understood as the ladder operators corresponding to the noise polarization ($\lambda = e$) and magnetization  ($\lambda = m$) excitations in the medium-assisted EM field, at frequency $\omega$, created or annihilated at position $\mb{r}$. The medium-assisted bosonic operators obey the canonical commutation relations \eqn{\sbkt{ \hat{\mb{f}}_{\lambda}\bkt{\mb{r}, \omega}, \hat{\mb{f}}_{\lambda'}\bkt{\mb{r}', \omega'} } = \sbkt{ \hat{\mb{f}}^{\dagger}_{\lambda}\bkt{\mb{r}, \omega}, \hat{\mb{f}}^\dagger_{\lambda'}\bkt{\mb{r}', \omega'} }=0,\\
\sbkt{ \hat{\mb{f}}_{\lambda}\bkt{\mb{r}, \omega}, \hat{\mb{f}}^\dagger_{\lambda'}\bkt{\mb{r}', \omega'} } = \delta_{\lambda\lambda'}\delta\bkt{\mb{r} - \mb{r}'}\delta\bkt{\omega - \omega'}.}

The electric and magnetic field operators evaluated  at the position of the particle are given as \eqn{\label{Era} \hat{\mb{E}}\bkt{\mb{r}_0}=& \sum_{ \lambda = e,m}\int \dd^3 r\int\dd\omega\non\\
& \sbkt{\dbar {G}_\lambda \bkt{\mb{r}_0, \mb{r}, \omega}\cdot \hat{\mb{f}}_{\lambda}\bkt{\mb{r},  \omega} + \text{H.c.}},\text{and}} \eqn{
\label{Bra}
\hat{\mb{B}}\bkt{\mb{r}_0}=& \sum_{ \lambda = e,m}\int \dd^3 r\int\dd\omega \,\non\\ &\left[\bkt{-\frac{i}{\omega}}\sbkt{\mb{\nabla}\times\dbar {G}_\lambda \bkt{\mb{r}_0, \mb{r}, \omega}}\cdot
 \hat{\mb{f}}_{\lambda}\bkt{\mb{r},  \omega} + \text{H.c.}\right]} respectively, where \eqn{[\vec{\bf \nabla}\times\bar{\bar {G}}_\lambda({\bf r},{\bf r}'\omega)]_{il} = \epsilon_{ijk}\partial_{r_j} [\bar{\bar{G}}_\lambda(\mb{r},\mb{r}',\omega)]_{kl}.} The coefficients $\bbar{G}_\lambda\pare{\mb{r}_1,\mb{r}_2,\omega}$  are defined as 
\eqn{\bbar{G}_e \pare{\mb{r},\mb{r}',\omega}=& i\frac{\omega^2}{c^2} \sqrt{\frac{\hbar}{\pi\epsilon_0}\Im[\epsilon \pare{\mb{r}',\omega}]} \bbar{G}\pare{\mb{r},\mb{r}',\omega}, \\
\bbar{G}_m \pare{\mb{r},\mb{r}',\omega}=& i\frac{\omega^2}{c^2} \sqrt{\frac{\hbar}{\pi\epsilon_0}\frac{\Im [\mu \pare{\mb{r}', \omega}]}{\abs{\mu\pare{\mb{r}',\omega}}^2}}\nabla\times \dbar{G}\pare{\mb{r},\mb{r}',\omega},}
with $\epsilon(\mb{r},\omega)$ and $\mu(\mb{r},\omega)$ as the space-dependent  permittivity and permeability, and $\bbar{G}\pare{\mb{r}_1,\mb{r}_2,\omega}$ as the Green's tensor for a point dipole near a planar semi-infinite surface \cite{Buh1,Buh2,GreenWelsch}.  Since we specifically want to study the effect of the presence of surface on the particle-field interaction, assuming that the corrections from the free space Green's tensor $\dbar G_0\bkt{\mb{r}_1,\mb{r}_2,\omega}$
such as the Lamb shift and the free space spontaneous emission are already taken into account, we consider only the part of the field that is scattered off the surface and interacts with the  dipole. This corresponds to  the scattering part of the total Green's tensor defined as
\eqn{\dbar G_{\text{sc}}\bkt{\mb{r}_1,\mb{r}_2,\omega} = \dbar G\bkt{\mb{r}_1,\mb{r}_2,\omega} - \dbar G_0\bkt{\mb{r}_1,\mb{r}_2,\omega}.}
We can thus define the interaction Hamiltonian corresponding to the interaction between the ED and MD with the part of the total EM field scattered off of the surface as
\eqn{\label{havsc}
H_{AV}^{\mr{sc}} = -\hat{\mb{d}}\cdot\hat{\mb{E}}_{{\mr{sc}}}\bkt{\mb{r}_0} -\hat{\mb{m}}\cdot\hat{\mb{B}}_{{\mr{sc}}}\bkt{\mb{r}_0},
}
where the field operators $\hat{\mb {E}}_{\mr{sc}}(\mb{r}_0)$ and $\hat{\mb {B}}_{\mr{sc}}(\mb{r}_0)$ are as defined in Eq.\,\eqref{Era} and Eq.\,\eqref{Bra}, with the total Green's tensor replaced by its scattering part  $(\dbar{G} (\mb{r}, \mb{r}',\omega)\rightarrow\dbar{G}_{\mr{sc}}(\mb{r}, \mb{r}',\omega))$.
\section{Scattering Green's tensor near a planar surface}
For a point dipole near an infinite half space, one can  write the scattering Green's tensor and its double curl  as \cite{Buh1}
\begin{widetext}
\eqn{\label{green}
 \dbar{G}_{\mr{sc}}\bkt{{\bf r}_0,{\bf r}_0,i\xi} &= \frac{1}{8\pi} \int_0^\infty\frac{\dd k_{\parallel}\,k_{\parallel} }{\kappa_\perp}e^{-2\kappa_\perp z_0}\left[-r_p\bkt{\kappa_\perp, i\xi}\frac{c^2}{\xi^2} \bkt{\begin{array}{ccc}
    \kappa_\perp^2 &0 &0 \\
    0 &\kappa_\perp^2 &0\\
    0& 0& 2k_\parallel^2
\end{array}}+r_s\bkt{\kappa_\perp, i\xi} \bkt{\begin{array}{ccc}
    1 &0 &0 \\
    0 &1 &0\\
    0& 0& 0
\end{array}} \right], \\
\label{greencc}\dbar{\mc{G}}_{\mr{sc}}\bkt{{\bf r}_0,{\bf r}_0,i\xi}&= \frac{\xi^2}{8\pi c^2} \int_0^\infty\frac{\dd k_{\parallel}\,k_{\parallel} }{\kappa_\perp}e^{-2\kappa_\perp z_0} \left[r_p\bkt{\kappa_\perp, i\xi} \bkt{\begin{array}{ccc}
    1 &0 &0 \\
    0 &1 &0\\
    0& 0& 0
\end{array}}-r_s\bkt{\kappa_\perp, i\xi}\frac{c^2}{\xi^2} \bkt{\begin{array}{ccc}
    \kappa_\perp^2 &0 &0 \\
    0 &\kappa_\perp^2 &0\\
    0& 0& 2k_\parallel^2
\end{array}} \right]
}
\end{widetext}

where $r_{s,p}$ are the Fresnel reflection coefficients for the $s$ and $p$ polarizations reflecting off the surface, and $\kappa_\perp^2 = \xi^2/c^2 +k_\parallel^2$. We see that all the information about the surface material is accounted for in the reflection coefficients which are given as
\eqn{\label{fresnel}
r_p\bkt{\kappa_\perp, i\xi} &=  \frac{\epsilon\bkt{i\xi}\kappa_\perp-\sqrt{\bkt{\epsilon\bkt{\mr{i}\xi}\mu\bkt{i\xi}-1}\xi^2/c^2+\kappa_\perp^2}}{\epsilon\bkt{i\xi}\kappa_\perp+\sqrt{\bkt{\epsilon\bkt{\mr{i}\xi}\mu\bkt{i\xi}-1}\xi^2/c^2+\kappa_\perp^2}},\non\\
r_s \bkt{\kappa_\perp, i\xi}&= \frac{\mu\bkt{i\xi}\kappa_\perp-\sqrt{\bkt{\epsilon\bkt{\mr{i}\xi}\mu\bkt{i\xi}-1}\xi^2/c^2+\kappa_\perp^2}}{\mu\bkt{i\xi}\kappa_\perp+\sqrt{\bkt{\epsilon\bkt{\mr{i}\xi}\mu\bkt{i\xi}-1}\xi^2/c^2+\kappa_\perp^2}}.
}
\section{Asymptotic potential for an particle near a metal surface}
\label{appB}
Let us consider a particle near a surface described by the Drude model with a dielectric permittivity given by Eq.\,\eqref{drude}. In the non-retarded limit, one can expand the Fresnel coefficients in Eq.\,\eqref{fresnel} to lowest order in $\sqrt{\epsilon(i\xi)-1}\xi/(\kappa_\perp c)$ as
\eqn{\label{rpapprox}
r_p\bkt{\kappa_\perp, i\xi}&\approx\frac{\epsilon\bkt{i\xi}-1}{\epsilon\bkt{i\xi}+1} - \frac{\epsilon\bkt{i\xi}\bkt{\epsilon\bkt{i\xi}-1}}{\bkt{\epsilon\bkt{i\xi}+1}^2}\frac{\xi^2}{\kappa_\perp^2c^2}\\
\label{rsapprox}
r_s\bkt{\kappa_\perp, i\xi}&\approx- \frac{1}{4}\bkt{\epsilon\bkt{i\xi}-1}\frac{\xi^2}{\kappa_\perp^2c^2}
}
We can then use the above to write the electric CP potential in the non-retarded limit $\bkt{\tilde z\ll1}$ as
\eqn{\label{denrMet}
  \tilde U_e^{\mr{NR}} \approx  -\frac{3\omega_p}{64 (\sqrt{2}\omega_e +\omega_p)\tilde{z}^3}.
}
In the retarded limit $\bkt{\tilde{z}\gg1}$ for the electric CP interaction, one can approximate all the response functions involved (particle polarizability and the Fresnel coefficients) by their static values $\bkt{\xi\rightarrow0}$, such that, one can then simplify the electric CP potential as 
\eqn{\label{deRMet}{  }\tilde U_e^{\mr{R}} \approx -\frac{3}{16\pi\tilde{z}^4},
}
which we can note from Table \ref{table1} is the same as the retarded potential for the particle near a perfectly conducting surface. This can be understood physically in terms of the fact that if the particle is far enough away from the surface such that the at the typical  frequencies relevant for the length scale involved the surface response effectively appears as that for a perfect conductor.

We use a similar approach to study the asymptotic behavior of the  level shifts due to ED and MD interactions for the particle in different regimes, as summarized  in Table \ref{table1}. The coefficients $\mc{C}_j^{k}$s for Drude and plasma models used in Table\,\ref{table1} are defined as follows
\eqn{     \label{ce3D}
\mc{C}_{e,D}^{(3)} &\equiv \frac{3\omega_p}{64 (\sqrt{2}\omega_e +\omega_p)},\\
\label{cm1D}
     \mc{C}_{m,D}^{(1)} &\equiv \frac{3\tilde{\omega}\eta S\omega_p}{64\omega_e} \left[\frac{\omega_p}{\omega_m+\omega_p/\sqrt{2}}\right.\non\\
     &\left.+\frac{\omega_p\bkt{\omega_m +2\gamma/\pi \ln (\gamma/\omega_m)}}{2\bkt{\omega_m^2 +\gamma^2}}\right],\\
     \label{cm1P}
     \mc{C}_{m,P}^{(1)} &\equiv \frac{3\tilde{\omega}\eta S\omega_p}{64\omega_e} \left[\frac{\omega_p}{\omega_m+\omega_p/\sqrt{2}}+\frac{\omega_p}{2\omega_m}\right]\non\\
     &+\frac{3}{64} \frac{\omega_p^2}{\omega_e^2} \eta S^2.}


\vspace{-0.2cm}\footnotesize{

}
\end{document}